\documentclass[12pt]{article}
\usepackage{graphicx}
\def\beq{\begin{equation}}
\def\eeq{\end{equation}}
\def\bea{\begin{eqnarray}}
\def\eea{\end{eqnarray}}

\def\roughly#1{\mathrel{\raise.3ex\hbox
{$#1$\kern-.75em\lower1ex\hbox{$\sim$}}}}

          \def\s{\tau}
     \def\t{\tilde}        
  \def\rr2{{1\over\sqrt{2}}}

\def\.{\!\cdot\!}    \def\:{\cdots}   \def\[{\left[}   \def\]{\right]}
\def\({\left(} \def\){\right)} 
\def\t3{{1\over\sqrt{3}}}
\def\s6{{1\over\sqrt{6}}}

\begin{document}
\begin{flushright}
UMISS-HEP-2008-03 \\
[10mm]
\end{flushright}

\begin{center}
\bigskip {\Large  \bf Deviation from Tri-Bimaximal Neutrino mixing
from the Charged Lepton Sector}  
\\[8mm]
Alakabha Datta\footnote{E-mail:
\texttt{datta@phy.olemiss.edu}}
\\[3mm]
\end{center}

\begin{center}
~~~~~~~~~~~ {\it Department  of Physics and Astronomy,}\\ 
~~~~~~~~~~~~{ \it University of Mississippi,}\\
~~~~~~~~~~~~{\it  Lewis Hall, University, MS, 38677.}\\
\end{center}

\begin{center} 
\bigskip (\today) \vskip0.5cm {\Large Abstract\\} \vskip3truemm
\parbox[t]{\textwidth} { The tri-bimaximal  mixing matrix is most likely 
the leading order term in the description of leptonic mixing. We derive deviations from tri-bimaximal  mixing(TBM)  due to corrections from the charged lepton sector. We assume a decoupled 2-3 symmetry, which is a $Z_2$ symmetry, in the charged lepton sector and another $Z_2$ symmetry in the neutrino sector. The TBM is obtained in the flavor symmetric limit. We consider
deviations from the TBM form arising from the  breaking of the 2-3 symmetry in the charged lepton sector but do not consider deviations  from the breaking of the flavor symmetry in the neutrino sector.
In particular we find the size of $\sin{\theta_{13}}$ to be related to the deviation $\sin{\theta_{12}}$ from $\t3$
while the deviation
of $\sin{\theta_{23}}$ from its tri-bimaximal value is $ \sim {m_{\mu} \over m_{\tau}}$.
 }
\end{center}

\thispagestyle{empty} \newpage \setcounter{page}{1}
\baselineskip=14pt

\section{Introduction}
We now know that neutrinos have masses and just like the quark mixing matrix there is a leptonic mixing matrix. This fact has been firmly established through
a variety of solar, atmospheric and terrestrial neutrino
oscillation experiments \cite{Bahcall:2004cc}.

The neutrino mixing, the PMNS matrix \cite{PMNS}, arise
from the lepton mass Lagrangian as follows:
\begin{eqnarray}
{\cal L}_m~=~\nu^T_\alpha C^{-1}{\cal M}_{\nu,\alpha\beta}\nu +
\bar{e}_{\alpha,L}M^e_{\alpha \beta}e_R + h.c.
\end{eqnarray}
We have assumed neutrinos to be Majorana particles. Diagonalizing the mass matrices by the transformations
$U^T_\nu{\cal M}_\nu U_\nu={\cal M}^\nu_{diag}$ and $U^{\dagger}_\ell
M^eV~ =M^e_{diag}$, one defines the neutrino mixing matrix as
$U_{PMNS}~=~U^{\dagger}_\ell U_\nu$. 
We will parametrize $U_{PMNS}$ as follows \cite{PDG}:
\begin{eqnarray}
 U_{PMNS}=\pmatrix{c_{12}c_{13} & s_{12}c_{13} &
s_{13} e^{-i\delta} \cr
-s_{12}c_{23}-c_{12}s_{23}s_{13}e^{i\delta}
&c_{12}c_{23}-s_{12}s_{23}s_{13}e^{i\delta} & s_{23}c_{13}\cr
s_{12}s_{23}-c_{12}c_{23}s_{13}e^{i\delta}
&-c_{12}s_{23}-c_{12}c_{23}s_{13}e^{i\delta} &
 c_{23}c_{13}}K,
 \label{PMNS}
\end{eqnarray} 
where $s_{13}=\sin \theta_{13}$, $c_{13}=\cos \theta_{13}$
with $\theta_{13}$ being the
reactor angle, $s_{12}=\sin \theta_{12}$, $c_{12}=\cos \theta_{12}$
with $\theta_{12}$ being the
solar angle, $s_{23}=\sin \theta_{23}$, $c_{23}=\cos \theta_{23}$
with $\theta_{23}$ being the
atmospheric angle, $\delta$ is the Dirac CP violating phase, and
$K~=~diag(1, e^{i\phi_1},e^{i\phi_2})$
 contains additional (Majorana)
CP violating phases $\phi_1, \phi_2$. We will ignore the
Majorana CP violating phases in this work.

Current experimental data is consistent with the tri-bimaximal mixing (TBM) form given below \cite{TBM}
\bea
 U_{MNS}^{sym} &= &
\pmatrix{
\sqrt{\frac{2}{3}}  & \frac{1}{\sqrt{3}} & 0 \cr
-\frac{1}{\sqrt{6}}  & \frac{1}{\sqrt{3}} & \frac{1}{\sqrt{2}} \cr
\frac{1}{\sqrt{6}}  & -\frac{1}{\sqrt{3}} & \frac{1}{\sqrt{2}}
}.\
\label{tbm}
\eea
In the literature various flavor symmetries have been considered to obtain the leptonic mixing and in particular TBM mixing \cite{Lam, others}. 
%
It is likely that the TBM form is the leading order term in the leptonic mixing matrix and the realistic mixing matrix is obtained by including small corrections to the TBM form.
Deviations from the TBM form have already been considered by several authors\cite{devs}. 
In this paper we assume that the TBM corresponds to a certain flavor symmetric limit in the charged lepton and neutrino sector. Realistic leptonic mixing is obtained by taking into account the violations of the flavor symmetries. We first identify the flavor symmetry in the charged lepton and the neutrino sector that generates the TBM form.
We identify the flavor symmetries to be invariance under $Z_2$ transformations. We then consider breaking of the flavor symmetry only in the charged lepton sector which in turn leads to the deviation of the leptonic mixing from the TBM form.
In this paper we will not consider deviations from the TBM mixing coming from the neutrino sector but we plan to consider them in a separate publication. Hence,
the fact that we do not consider deviations to the TBM from the neutrino sector in this paper does not mean that we believe there are no deviations from TBM mixing from the neutrino sector.
Contributions from the charged lepton sector to leptonic mixing have been considered previously\cite{CCC} and the $Z_2$ symmetry in leptonic mixing have been discussed before in Ref.~\cite{Z2} but we believe the results presented in the paper are new.

As indicated earlier, there is a great deal of work on model building of mixing matrices in the lepton sector. In this paper we are less sophisticated in model building and try to arrive at the flavor symmetry for TBM purely from phenomenological considerations. One of the key inputs in our consideration is the fact that in the flavor symmetric limit we  require that the charged lepton and neutrino mass matrices to be diagonalized by matrices composed of pure numbers and independent of the parameters of the mass matrices.
This requirement  puts important constraints on the structure of the charged lepton and neutrino mass matrices. 

To begin our analysis we start from the fact that the
 quark and charged leptons exhibit similar hierarchical structures. We  therefore assume the same flavor structure for them. We  also use similar parametrization to represent deviation from the flavor symmetric limit for the quark and the charged lepton sector. For the flavor symmetry in the charged lepton sector we  use the decoupled 2-3 flavor symmetry\cite{model23}. In the quark and the charged lepton sector this symmetry gives an understanding of the mass splitting between the second and the third generations.
 The first generation is decoupled from the second and the third in the flavor symmetric limit. The decoupled 2-3 symmetry arises if we require the general 2-3 symmetric mass matrix be diagonalized by a matrix of pure numbers. The decoupled 2-3 symmetric form is invariant under a $Z_2$ transformation. After fixing the structure of the charged lepton mass matrix in the flavor symmetric limit we  fix the structure of the neutrino mass matrix by requiring that we obtain TBM mixing in the flavor symmetric limit. The resulting neutrino mass matrix is found to be invariant under another $Z_2$ transformation. 
 
 We  next consider possible deviations from the TBM structure coming from the symmetry breaking in the charged lepton sector. The parametrization
 of flavor symmetry breaking in the charged lepton sector is taken to be similar in structure to the parametrization of symmetry breaking in the quark sector.
 Our parametrization is similar to the one considered in Ref~\cite{fxing} for the quark mass matrices. 
Finally, assume CP conservation in our analysis.

The paper is organized in the following manner:
We begin in Sec. 2  with a discussion of the flavor symmetric limit
that leads to the TBM mixing. In 
Sec. 3 we study the effect of symmetry breaking from the charged lepton sector to generate the realistic leptonic mixing matrix and finally in Sec. 4 we
conclude with a summary of the results reported in this work.

\section{ The tri-bimaximal mixing from flavor symmetry}
The tri-bimaximal mixing matrix is composed of pure numbers. It is natural to assume that the neutrino and charged lepton components of the mixing matrix are also composed of pure numbers. For the charged lepton sector we assume that the flavor symmetry is a decoupled 2-3 symmetry\cite{model23}. This symmetry is just a special case of the 2-3 or the $\mu-\tau$ symmetry \cite{mutau} which has been widely studied in the literature.
In the 2-3 symmetric limit the Yukawa matrix of the charged lepton, $Y^L$, has the following structure,
\bea
Y^L & = &\pmatrix{l_{11} & l_{12}
&  -l_{12}  \cr  l_{12}  &  l_{22}  &  l_{23}\cr  -l_{12}  &l_{23}  &l_{22}
}. \
\eea
This matrix is diagonalized as
\bea
U^{\dagger} Y^L U & = & Y^L_{diag}, \nonumber\\
U & = &\pmatrix{1 & 0
&  0  \cr  0  & 1/\sqrt{2}   &  1/\sqrt{2}\cr  0  &- 1/\sqrt{2}  &1/\sqrt{2}
} 
\cdot
\pmatrix{\cos{\theta} & \sin{\theta}
&  0  \cr  -\sin{\theta}  & \cos{\theta}   &  0\cr  0  &0  &1
},\ 
\label{lam}
\eea
where the mixing angle $\theta$ is determined by the positive solution to
 \bea
 \tan \theta  & = & {2 \sqrt{2}l_{12} \over {
 l_{22}-l_{23}-l_{11} \pm 
 \sqrt{ (l_{22}-l_{23}-l_{11})^2+8l_{12}^2}
 }
 }. \
 \eea
 The eigenvalues of $Y^L$ are ${ 1 \over 2} [l_{11}+l_{22}-l_{23} \pm \sqrt{(l_{11}-l_{22}+l_{23})^2+8l_{12}^2} ]$ and $l_{22}+l_{23}$.

According to our assumption, the elements of the matrix that diagonalizes $Y^L$ must be pure numbers. It is clear that we can achieve that
 by setting $l_{12}=0$( $\theta =0$). We are going to consider this limit of the 2-3 symmetry, which we will call the decoupled 2-3 symmetry, as the flavor symmetry in the charged lepton sector sector. In this decoupled 2-3 symmetric limit \cite{model23} the first generation is decoupled from the second and third generations.
 
Let us represent the Yukawa matrix with the decoupled 2-3 symmetry by $Y^L_{23}$. We will take the Yukawa matrix to have the form, 
\bea
Y^L_{23} & = &
\pmatrix{l_{11} & 0
&  0  \cr  0  &  \frac{1}{2}{l_{22}}  &  \frac{1}{2}{l_{23}}\cr  0  
&\frac{1}{2}{l_{23}}  &\frac{1}{2}{l_{22}}}. \
 \label{23sym}
 \eea
This Yukawa matrix  $Y^L_{23}$ is diagonalized by the unitary matrix $W^{l}_{23}$ given by,
\bea
W^l_{23} &  = & \pmatrix{1 & 0
&   0  \cr   0   &  -\frac{1}{\sqrt{2}}   &  \frac{1}{\sqrt{2}}\cr   0
&\frac{1}{\sqrt{2}} &  \frac{1}{\sqrt{2}}}.\
\label{wl} 
\eea 
Note that this matrix differs from the one in Eq.~\ref{lam} in the limit $\theta=0$ by an irrelevant diagonal phase matrix.

Writing the diagonalized Yukawa matrix as $Y^{L}_{23d}$  we have,
\bea 
Y^L_{23d} & = & W^{l \dagger}_{23} Y^L_{23} W^l_{23} 
=\pmatrix{l_{11} & 0
&  0  \cr  0  &   \frac{1}{2}{(l_{22}-l_{23})}  &  0\cr  0  &0  &
\frac{1}{2}{(l_{22}+l_{23})}}. \  
\label{diag23sym} 
\eea 

It is interesting to consider the underlying flavor symmetry  of $Y^L_{23}$ in Eq.~\ref{23sym}. One can easily check that this matrix satisfies
\bea
G_l^TY^L_{23}G_l & = & Y_L^{23}, \
\label{z2lepton}
\eea
where
\bea
G_l^T& = & G_l=\pmatrix{1 & 0
&  0  \cr  0  &  0 &  1\cr  0  &1  &
0} \nonumber\\
G_l^2& = & 1\
\label{glepton}
\eea
It is easy to recognize $G_l$ to be a representation of the $Z_2$ symmetry group. In other words $Z_2= ( I, G_l) $ where $I$ is the $ 3 \times 3$ identity matrix.

Having established the flavor symmetry in the charged lepton sector, we use the TBM mixing as an input to identify the flavor symmetry in the neutrino sector. The TBM mixing is a specially case of mixing matrix with 2-3 flavor symmetry \cite{mutau}. In the 2-3 symmetry limit 
the PMNS matrix, with $s_{13}=0$, is 
given by,
\bea
U_{PMNS}^{s} & = & \pmatrix{c_{12} &s_{12}&0\cr
-\rr2 s_{12}
&\rr2 c_{12}  & \rr2\cr
\rr2 s_{12}
&- \rr2 c_{12} &
 \rr2}.
\label{mns23}
\eea
The TBM form is obtained by setting $s_{12}= { 1 \over {\sqrt{3}}}$.
We can then express $U_{PMNS}^{s}$, as,
\bea
U_{PMNS}^{s}&= U^{\dagger}_\ell U_\nu, \
\label{nulepton}
\eea
where
\bea
U_\ell & = & W^l_{23}, \nonumber\\
U_\nu & = & \pmatrix{c_{12}&-s_{12}&0\cr s_{12}&c_{12}&0\cr 0&0&1\cr}\pmatrix{1&0&0\cr 0& -1&0\cr 0&0&1\cr},\
\label{tbmfactor} 
\eea 
with 
\bea
s_{12}&= & \sin{\theta_{12}}= { 1 \over {\sqrt{3}}} \nonumber\\
c_{12}& = & \cos{\theta_{12}}= \sqrt{{2 \over 3}}.\
\label{solar}
\eea
 Hence in the flavor symmetric limit,
 the neutrino matrix, $U_{\nu}$ is just a combination of a simple rotation matrix and a 
phase matrix. 

Let us  discuss the structure of the neutrino matrix in the flavor symmetric limit. It can be easily seen that ${\cal {M}}_\nu$ is given as
\bea
{\cal {M}}_\nu & = &
\pmatrix{
a  & \sqrt{2}(a-b) & 0 \cr
\sqrt{2}(a-b)  & b & 0 \cr
0  & 0 & c
},\
\label{neu_mass}
\eea
with $U^T_\nu{\cal M}_\nu U_\nu={\cal M}^\nu_{diag}$. We see that the neutrino mass matrix exhibits decoupling of the first two generations from the third generation.
Let us assume that ${\cal {M}}_\nu$ has the general structure
\bea
{\cal {M}}_\nu & = &
\pmatrix{
a  & d & 0 \cr
d  & b & 0 \cr
0  & 0 & c
},\
\eea
Now we want the matrix that diagonalizes the mass matrix above to be composed of pure numbers independent of the values of the parameters in ${\cal {M}}_\nu $ . It is easily checked that ${\cal {M}}_\nu $  is diagonalized by $U_{\nu}$ given in Eq.~\ref{tbmfactor} where for $\theta_{12}$ in Eq.~\ref{solar} we have
\bea
\tan{2 \theta_{12}} & = & \frac{2d}{(a-b)},\
\eea
Hence for $\theta_{12}$ to be independent of the parameters $a,b$ and $d$
we should have
\bea
d &=& k(a-b),\
\eea
where $k$ is a number. Hence,
\bea
{\cal {M}}_\nu & = &
\pmatrix{
a  & k(a-b) & 0 \cr
k(a-b)  & b & 0 \cr
0  & 0 & c
},\
\label{neusym}
\eea
The form for ${\cal {M}}_{\nu} $ in Eq.~\ref{neu_mass} is obtained with
 $k= \sqrt{2}$.

It is again interesting to consider the underlying flavor symmetry  of 
${\cal {M}}_\nu$  in Eq.~\ref{neu_mass}. One can easily check that this matrix satisfies
\bea
G_{\nu}^T{\cal {M}}_{\nu} G_{\nu} & = {\cal {M}}_{\nu},\
\label{z2neutrino}
\eea
where
\bea
G_{\nu}^T& = & G_{\nu}=\pmatrix{\frac{1}{3} & \frac{2 \sqrt{2}}{3}
&  0  \cr  \frac{2 \sqrt{2}}{3}  &  -\frac{1}{3} &  0\cr  0  &0  &
1} \nonumber\\
G_{\nu}^2& = & I,\
\label{gneutrino}
\eea
with $I$ being  $ 3 \times 3$ identity matrix.
It is again easy to recognize $G_{\nu}$ to be a representation of the $Z_2$ symmetry group. In other words $Z_2= ( I, G_{\nu} )$.

 The breaking  of the decoupled 2-3 symmetry or equivalently the $Z_2$ symmetry in the charged lepton sector will cause deviation from the TBM form and we
study this deviation in the next section. As indicated in the introduction, in this paper we will not consider deviations from the TBM mixing coming from the neutrino sector. 

 \section{Symmetry Breaking}
 We now consider the breaking of the decoupled 2-3 symmetry in the charged lepton sector. The diagonal charged lepton mass matrix is, 
 \bea
 M^L_{diag} & = &\pmatrix{\pm m_e & 0
&  0  \cr  0  &  \pm m_{\mu}  &  0 \cr  0  & 0  & \pm m_{\tau}
} =
 \frac{v}{\sqrt{2}}\pmatrix{l_e & 0
&  0  \cr  0  &  l_{\mu}  &  0 \cr  0  & 0  & l_{\tau}
}, \
\label{mass_yukawa}
\eea 
where $v$ is the v.e.v of the higgs field.
The diagonalized mass matrix  $M^L$ in the flavor symmetric limit is  given by,
\bea 
M^D_{diag} & = & W^{l \dagger}_{23} M^L W^l_{23} =\pmatrix{\frac{v}{\sqrt{2}}l_{11} & 0
&  0  \cr  0  &   \frac{v}{\sqrt{2}}\frac{1}{2}{(l_{22}-l_{23})}  &  0\cr  0  &0  &
\frac{v}{\sqrt{2}}\frac{1}{2}{(l_{22}+l_{23})}}. \
\label{mass} 
\eea 
The  charged lepton  masses   are  given   by,
  \bea  m_e   &  =   &  \pm
\frac{v}{\sqrt{2}}l_{11},       \nonumber\\
  m_{\mu}       &=&      \pm
\frac{v}{\sqrt{2}}{(l_{22}-l_{23}) \over 2},   \nonumber\\
  m_{\tau}   &  =   & \pm   
 \frac{v}{\sqrt{2}}{(l_{22}+l_{23}) \over 2}.\ 
\eea
Since $m_{\mu} <<  m_{\tau}$ there has to be a  fine tuned cancellation between
$l_{22}$ and $l_{23}$ to produce  the muon mass. Hence, it is
more  natural to  consider  the symmetry  limit $l_{22}=l_{23}$  which
leads to  $m_{\mu}=0$. The muon  mass is then generated due to symmetry breaking. We, therefore, consider the structure,
\begin{eqnarray}
 Y^L_{23} &= &\pmatrix{l_e & 0 &  0 \cr 0 & \frac{1}{2}{l_{T}}(1+2\chi_l) & \frac{1}{2}{l_{T} }\cr 0
&\frac{1}{2}{l_{T}} & \frac{1}{2}{l_{T}}}.\
 \label{23dsymbreak1}
 \eea
 The structure above breaks the the $Z_2$ symmetry in Eq.~\ref{z2lepton}.
Note that we  do not
break  the $2-3$ symmetry  in the  $23$ element  so that  the Yukawa
matrix remains  symmetric. The matrix $Y^L_{23}$  is now diagonalized
by the unitary matrix, $U_l= W^l_{23} R^l_{23}$ where
\bea
R^l_{23} & = & \pmatrix{1 & 0
&  0  \cr  0  &  c_{23l}   &  s_{23l}\cr  0  &-s_{23l}  &
c_{23l}}, \nonumber\\ 
c_{23l} & = & \cos { \theta_{23l}} ; s_{23l}  =  \sin { \theta_{23l}}. \
\label{23d}
 \eea
We can write $Y^L_{23}$ as
\bea
Y^L_{23} & = & U_l \pmatrix{l_e & 0
&  0  \cr  0  &  l_{\mu}  &  0 \cr  0  & 0  & l_{\tau}
}U_l^{\dagger}, \
\label{another}
\eea
 where $l_{e,\mu,\tau}$ are the diagonal Yukawa couplings defined in Eq.~\ref{mass_yukawa}.
One can then obtain 
 the nonzero elements of $Y^L_{23}$ as,
\bea
(Y^{L}_{23})_{11} & = & l_e, \nonumber\\
(Y^{L}_{23})_{22} & = & \frac{l_{\mu}+l_{\tau}}{2} -(l_{\tau}-l_{\mu})s_{23l}c_{23l}, \nonumber\\
(Y^{D}_{23})_{23} & = &\frac{l_{\tau}-l_{\mu}}{2} -(l_{\tau}-l_{\mu})s_{23l}^2=(Y^{D}_{23})_{32}, \nonumber\\
(Y^{D}_{23})_{33} & = & \frac{l_{\mu}+l_{\tau}}{2} +(l_{\tau}-l_{\mu})s_{23l}c_{23l} .\ 
\eea
As the Yukawa matrix must have the form in Eq.~\ref{23dsymbreak1} we have
 $(Y^{L}_{23})_{23}=(Y^{L}_{23})_{33}$ which then leads to
 \bea
 \tan{\theta_{23l}} & = & \frac{1}{2} \left[ z_l-1 + \sqrt{z_l^2-6z_l+1} \right], \
 \label{theta23solnd}
 \eea
 where $z_l={ l_{\mu} \over l_{\tau}} =\pm{ m_{\mu} \over m_{\tau}}$ and we have chosen the solution that leads to small angle
 $\theta_{23l}$
 and to small flavor symmetry breaking.
 Keeping terms to first order in $z_l$ we get
 \bea
 \tan{\theta_{23l}} & \approx  & -z_l. \
 \label{theta23solnapproxd}
 \eea
  We further obtain for $\chi_l$ and $l_T$ in Eq.~\ref{23dsymbreak1},
 \bea
 \chi_l &=& - \tan{2 \theta_{23l}} \approx 2 z_l,\nonumber\\
 l_{T} & = & (l_{\tau}-l_{\mu})\cos{ 2 \theta_{23l}}.\
 \label{chi}
 \eea
  
 To obtain a realistic PMNS matrix, we take into account the mixing involving the first and the second generation.
 We will assume that the Yukawa matrix $Y^{L}$ is now diagonalized by the unitary matrix $U_l$ given by, 
 \bea
 U_l & = & W^l_{23}R^{l}_{23}R^{l}_{12} , \
 \label{diagmat}
 \eea
 where
 \bea
 R_{12l} & = & \pmatrix{c_{12l} & s_{12l}
&  0  \cr  -s_{12l}  &  c_{12l}   &  0\cr 0  &
0& 1}, \nonumber\\
c_{12l} & = & \cos { \theta_{12l}} ; s_{12l}  =  \sin { \theta_{12l}},\
\label{12l}
\eea
and $R^{l}_{23}$ is given by Eq.~\ref{23d}. The Yukawa matrix $Y^L$ reduces to $Y^L_{23}$ in Eq.~\ref{23dsymbreak1} when $\theta_{12l}=0$.
 
This scheme of mass matrices was considered in Ref~\cite{fxing} for the quark mass matrices. Our scheme is equivalent to the one in Ref~\cite{fxing} up to a unitary transformation by $W^l_{23}$. 

We can now consider deviations from the TBM mixing by including the symmetry breaking. The leptonic mixing matrix is now given by
\bea
U_{PMNS}&= U^{\dagger}_\ell U_\nu, \
\label{nuleptonfinal}
\eea
where
$U_\ell  =  W^l_{23} R^l_{23}R^l_{12}$ and $U_{\nu}$ is given in Eq.~\ref{tbmfactor}.
We then obtain for the elements of the leptonic mixing,
\bea
U_{11} & = &\sqrt{\frac{2}{3}}\left[c_{12l}+\frac{1}{2}( s_{12l}c_{23l}+ s_{12l}s_{23l})\right],\nonumber\\
U_{12} & = & 
\t3 \left[c_{12l}- s_{12l}c_{23l}- s_{12l}s_{23l}\right],\nonumber\\
U_{13} & = & -\rr2 s_{12l}
(c_{23l}- s_{23l}), \nonumber\\
U_{21} & = & 
-\s6(c_{12l}c_{23l} + c_{12l}s_{23l}-2s_{12l} ), \nonumber\\
U_{22} & = &
\t3(c_{12l}c_{23l} + c_{12l}s_{23l}+s_{12l} ), \nonumber\\
U_{23} & = &
\rr2 c_{12l}(c_{23l}-s_{23l}), \nonumber\\
U_{31} & = &
\s6(c_{23l}- s_{23l}), \nonumber\\
U_{32} & = & -\t3(c_{23l}- s_{23l}), \nonumber\\
U_{33} & = & \rr2(c_{23l}+ s_{23l}).\
\label{realmix}
\eea
{}Following Ref.~\cite{king} we  expand the angles in Eqn.~\ref{PMNS}
as
\begin{equation}
s_{13} = \frac{r}{\sqrt{2}}, \ \ s_{12} = \frac{1}{\sqrt{3}}(1+s),
\ \ s_{23} = \frac{1}{\sqrt{2}}(1+a),
\label{rsa}
\end{equation}
where the three real parameters
$r,s,a$  describe the deviations of the reactor, solar and
atmospheric angles from their tri-bimaximal values. As in 
Ref.~\cite{king} we  use
global fits of the conventional mixing angles
\cite{Maltoni:2004ei} that can be translated into the $2\sigma$ ranges
\begin{equation}
0<r<0.22, \ -0.11<s<0.04, \ -0.12<a<0.13.
\end{equation}
To first order in
in $r,s,a$ the lepton mixing matrix can be written as,
\begin{eqnarray}
U \approx
\left( \begin{array}{ccc}
\sqrt{\frac{2}{3}}(1-\frac{1}{2}s)  & \frac{1}{\sqrt{3}}(1+s) & \frac{1}{\sqrt{2}}re^{-i\delta } \\
-\frac{1}{\sqrt{6}}(1+s-a + re^{i\delta })  & \frac{1}{\sqrt{3}}(1-\frac{1}{2}s-a- \frac{1}{2}re^{i\delta })
& \frac{1}{\sqrt{2}}(1+a) \\
\frac{1}{\sqrt{6}}(1+s+a- re^{i\delta })  & -\frac{1}{\sqrt{3}}(1-\frac{1}{2}s+a+ \frac{1}{2}re^{i\delta })
 & \frac{1}{\sqrt{2}}(1-a)
\end{array}
\right).
\label{MNS1}
\end{eqnarray}
Comparing with Eq.~\ref{realmix} and expanding to first order in $s_{12l}$ and $s_{23l}$ we find, taking $\delta= \pi$,
\bea
s & \approx & -  s_{12l},  \nonumber\\
r & \approx & -s, \nonumber\\
a & \approx & -s_{23l} \approx  \mp \frac{m_{\mu}}{m_{\tau}}.\
\label{relations}
\eea
Since the present data prefers a negative value for $s$ \cite{king} and $r$ is positive, we have chosen $\delta= \pi$.
We therefore find that,
$s_{13}$ is given by the deviation of $s_{12}$ from $\t3$ while the deviation
of $s_{23}$ from its tribimaximal value is $ \sim {m_{\mu} \over m_{\tau}}$.
In our discussion we do not consider CP violation.

\section{ Conclusion}
In conclusion in this paper we have considered deviations from the tri-bimaximal mixing due to corrections from the charged lepton sector.
The TBM leptonic mixing is obtained in the limit of a decoupled 2-3 symmetry, which is a $Z_2$ symmetry, in the charged lepton sector and another $Z_2$ symmetry in the neutrino sector. This symmetry is broken and inclusion of the symmetry breaking in the charged lepton sector leads to deviations from the TBM form. We ignored corrections to TBM from the neutrino sector which will be considered in a separate publication.  In particular we found the size to $\sin{\theta_{13}}$ to be related to the deviation of $\sin{\theta_{12}}$ from $\t3$
while the deviation
of $s_{23}$ from its tribimaximal value was found to be $ \sim {m_{\mu} \over m_{\tau}}$.



\end{document}